\documentclass[sigconf]{acmart}

\usepackage{booktabs} 
\usepackage{multirow}
\usepackage{rotating}

\setcopyright{rightsretained}

\newtheorem{prop}{Proposition}
\copyrightyear{2019}
\acmYear{2019} 
\setcopyright{iw3c2w3}
\acmConference[WWW '19]{Proceedings of the 2019 World Wide Web Conference}{May 13--17, 2019}{San Francisco, CA, USA}
\acmBooktitle{Proceedings of the 2019 World Wide Web Conference (WWW '19), May 13--17, 2019, San Francisco, CA, USA}
\acmPrice{}
\acmDOI{10.1145/3308558.3313693}
\acmISBN{978-1-4503-6674-8/19/05}

\begin{document}
\title{The Few-get-richer: A Surprising Consequence of Popularity-based Rankings}
\titlenote{F. Germano acknowledges financial support from Grant ECO2017-89240-P (AEI/FEDER, UE) from the Spanish MINECO. 
G. Le Mens benefited from financial support from Ramon y Cajal Fellowship RYC-2014-15035 and Grant \#AEI/FEDER UE-PSI2016-75353 from the Spanish MINECO, and ERC Consolidator Grant \#772268 from the European Commission.
Germano and Le Mens also acknowledge financial support from the Spanish MINECO through the Severo Ochoa Program for Centers of Excellence in R\&D, Grant SEV-2015-0563.
V. G\'omez is funded by the Ramon y Cajal program RYC-2015-18878 (AEI/MINECO/FSE,UE) and the Mar\'{i}a de Maeztu Units of Excellence Programme (MDM-2015-0502). The research leading to these results has received funding from ``La Caixa" Banking Foundation.
Data from the experiment are available at \href{https://osf.io/nwjyf/}{https://osf.io/nwjyf/}.}

\author{Fabrizio Germano}
\affiliation{%
 \institution{U. Pompeu Fabra, BGSE}
}
\email{fabrizio.germano@upf.edu}

\author{Vicen\c{c} G\'omez}
\affiliation{%
 \institution{U. Pompeu Fabra}}
\email{vicen.gomez@upf.edu}
\author{Ga\"el Le Mens}
\affiliation{%
 \institution{U. Pompeu Fabra, BGSE, SDU}}
\email{gael.le-mens@upf.edu}


\begin{abstract}
Ranking algorithms play a crucial role in online platforms ranging from  search engines to recommender systems. In this paper, we identify a surprising consequence of popularity-based rankings: the fewer the items reporting a given signal, the higher the share of the overall traffic they collectively attract. This few-get-richer effect emerges in settings where there are few distinct classes of items (e.g., left-leaning news sources versus right-leaning news sources), and items are ranked based on their popularity.  We demonstrate analytically that the few-get-richer effect emerges when people tend to click on top-ranked items and have heterogeneous preferences for the classes of items. Using simulations, we analyze how the strength of the effect changes with assumptions about the setting and human behavior. We also test our predictions experimentally in an online experiment with human participants. Our findings have important implications to understand the spread of misinformation.
\end{abstract}

%
%



\maketitle

\section{Introduction}
Ranking systems are at the core of many online services, including search engines, recommender systems, or news feeds in social media. Recent research suggests that the underlying ranking algorithms may impact society, playing an active role in the spread of misinformation~\cite{vosoughi2018spread}, political polarization~\cite{Dandekar5791}, or trustworthiness~\cite{diaz2008through}. They might also reinforce existing judgment biases~\cite{Baeza18}.

Rankings systematically affect the information people access about products, services, events, or ideas, because users are more likely to click on top-ranked items~\cite{joachims2017accurately, wang2018position, LeMens2018WisdomOfTheCrowd}. When items are ranked based on popularity, this leads to a self-reinforcing dynamics according to which  popular items become increasingly more popular~\cite{salganik2006experimental}.

In this paper, we identify a surprising effect of popularity-based rankings.
Consider a setting with two distinct classes of news sources that differ in their political orientations, e.g., left-leaning or right-leaning. We show that, under a fairly broad set of conditions, \emph{the total share} of web traffic (proportion of clicks) attracted by a given class of  news sources \emph{decreases} with the number of news sources in that class. We call this phenomenon the `few-get-richer' effect. For example, if there are 20 news sources, the total number of clicks on left-leaning sources will be larger when there are just 3 of these sources than when there are 17 of them. 

Intuition suggests that popular items should be more relevant and trustworthy than unpopular ones. Yet, extensive research indicates that popularity is often not very informative about quality, especially in settings characterized by `rich-get-richer' dynamics (sometimes called the `Matthew effect')~\cite{LeMensManSci2017, LeMensTopicsInCS2018,merton1968matthew,salganik2006experimental},
or information cascades~\cite{banerjee1992simple,bikhchandani1992theory}. In these settings, the randomness inherent to the dynamics of the system implies that items that become the most popular are not always those with the best quality. The `few-get-richer' effect adds to research on the `rich-get-richer' dynamics by showing that popularity-based rankings do not only create `noise' in the ranking, but can also lead to a systematic ranking bias: when there are two distinct classes of items, items from the smaller class become better ranked than similar items from the larger class.

The few-get-richer effect emerges in settings characterized by two design features. The first feature consists in the ranking of items in terms of popularity (i.e., items with more clicks are higher ranked). The second feature is a partition of the available items in two (or more) distinct classes. 

We make two reasonable behavioral assumptions. The first assumption is users' tendency to click on top-ranked items. The second assumption is that users have heterogeneous preferences for the item classes. Some users have a preference for items of a particular class, while others have a preference for items of other classes. Still other users are indifferent to the item class.

Returning to our news search example, suppose there are few left-leaning and many right-leaning news sources. We assume there are three types of users: left-leaning, right-leaning, and indifferent. The heterogeneous preference assumption means that left-leaning users are more likely to click on left-leaning news sources, right-leaning users are more likely to click on right-leaning news sources, and indifferent users click exclusively based on rank. Even if the left-leaning news sources are unpopular, left-leaning individuals will seek them out. Because there are few such news sources, the clicks of these left-leaning individuals will be concentrated on a few news sources, and these sources will tend to `shoot up to the top'.
Once a news source has gotten close to the top, it will attract not only the clicks from the left-leaning individuals, but also the clicks of indifferent users, simply because of the rich-get-richer dynamics. This is the few-get-richer effect.

\section{Related Work}

Our results contribute to the understanding of the limitations of recommender systems~\cite{herlocker2004evaluating,mcnee2006being,LeMens2018WisdomOfTheCrowd,ciampaglia2018how,Ilan2014, Che2018}, with direct applications to the design of fair, transparent and efficient ranking systems~\cite{Castillo,Biega,zehlike2017fa}, as well as methods to reduce the spread of  misinformation~\cite{kim2018leveraging,vosoughi2018spread} or uncivil behavior~\cite{muddiman2017news,cheng2017anyone}.

An extensive literature on modeling (click) user behavior is weakly related to our work~\cite{zhang2011user,chapelle2009dynamic,dupret2008user,richardson2007predicting}. Closer to our work, several papers have proposed models of the dynamics of interactions between individual searches and ranking algorithms, e.g., for understanding the feedback loop between ranking system and user queries~\cite{demange}, explaining the observed mitigation of search engines' popularity bias~\cite{Fortunato12684}, or the competition of memes using limited attention~\cite{weng2012competition}. 
The paper closest to ours is~\cite{Germano2018}, which also obtains a few-get-richer effect in a model 
where individuals get multiple signals and where (news) items are ranked via a probabilistic popularity-based ranking. Besides being simpler, our model works with a discrete and deterministic ranking of the websites rather than a continuous and probabilistic one. Among other things, this allows for a tighter connection with the experiment of Section~\ref{sec:exp}.

\section{The Model}
\label{sec:model}

We present a stylized model of a search environment where individuals use a search engine to look for information on a binary issue.
At the center of the model is the ranking algorithm, which ranks and directs individuals to the different websites, based on the popularity of individuals' choices. 


\subsection{Model of the Search Environment} There are $M$ items, each of which belongs to exactly one of two classes.
For example, the items can be news articles and the class of an article can be whether the source is known to be left-leaning or right-leaning.
A different, visual, example can comprise images about animals, some of which are cats while the others are dogs. 
In general, each of the $M$ items is characterized by a binary signal~$\{ 0, 1 \}$ that defines its class;
let $M_k$ denote the set and number of items of class $k$, $k\in \{ 0, 1 \}$. 

There is a single popularity-based ranking algorithm which, starting with a given initial ranking $(r_1)$, ranks all the items in $M$ according to the number of clicks received. Let $r_n \in \{ 1, \ldots , M \}^M$ denote the ranking seen by individual $n$, where $r_{n,m} \in \{ 1, \ldots, M \}$ denotes the rank of item $m$ observed by individual $n$. 

There are $N$ individuals, each of which is characterized by a parameter $\gamma_n$, which can be of one of three types. The three types are denoted by Type $0$, Type $1$ and Type $2$, and their proportions in the population are $p_0$, $p_1$ and $p_2=1-p_0-p_1$, respectively.  
Type 0 (resp.~Type 1) individuals have a preference for clicking on items of class $0$ (resp. class $1$).  Type $2$ individuals are indifferent between clicking on items of class $0$ or of class $1$ {\em absent ranking}. We represent the set of possible types by $\Gamma=\{\Gamma_0, \Gamma_1, \Gamma_2\}$, where $\Gamma_i$ denotes the probability that an individual of Type $i$ (i.e., with $\gamma_n= \Gamma_i$) clicks on an item of class $0$ {\em absent ranking}. Our assumptions about the preferences of the individuals in the three types imply $\Gamma_0>\frac{1}{2}$, $\Gamma_1<\frac{1}{2}$ and $\Gamma_2=\frac{1}{2}$. 


We summarize these preferences in terms of propensities $\varphi_{n,m}$ with which individual $n$ with $\gamma_n\in\Gamma$ clicks on item $m$, defined by:
\begin{align}\label{eq:phi}
\varphi_{n,m} & = 
\begin{cases}
\vspace{.08in}
\frac{\gamma_n}{M_0}     & \text{ if $m \in M_0$} \\ 
\frac{1-\gamma_n}{M_1} & \text{ if $m \in M_1$} .\\
\end{cases}
\end{align}
For simplicity, in this section, we consider the extreme and symmetric case, where $\Gamma_0=1$, $\Gamma_1=0$ and $p_0=p_1=p>0$.

\subsection{Model of Stochastic Choices}
Individuals enter one after the other and observe the ranked list of items, and based on the ranking and on their preference for the signals given by $\varphi_{n,m}$, they click on one of the $M$ items according to a probabilistic choice function obtained as follows. 
We use the function $\beta^{\left(M-r_{n,m}\right)}$ to weigh the propensities $\varphi_{n,m}$, whereby~$\beta>~1$ calibrates an individual's search cost or attention bias, so that an item ranked exactly one position higher has $\beta$ times as much probability of being clicked. The probability individual $n$ clicks on website $m$ is given by:

\begin{equation}\label{eq:rho}
\rho_{n,m} = 
\frac{\beta^{\left(M-r_{n,m}\right)} \varphi_{n,m}}{\sum_{m' \in M} \beta^{\left(M-r_{n,m'}\right)} \varphi_{n,m'}} .
\end{equation}
Thus, individual $n$ observes $r_{n,m}$ and clicks  on a website according to $\rho_{n,m}$. His click gets recorded by the ranking algorithm, which updates its ranking to $r_{n+1}$. This affects the ranking of the websites observed by the next individual, who clicks on a website according to $\rho_{n+1,m}$ and so on. We denote the total clicking probability on items of class $k$ ($M_k$) by $\rho_{n,M_k}=\sum_{m \in M_k}\rho_{n,m}$.

Overall, this defines a search environment $\mathcal{E}$ with parameters $(M, M_1); (N, \beta, \Gamma, p); r_1$.
We assume $\beta > 1$ and $0 < p < \frac{1}{2}$ so that individuals are subject to an attention bias and are heterogeneous in the sense that there are nonzero shares of Types 0, 1 and 2.

\subsection{The Few-Get-Richer Effect}
The following result shows how few items of a given class can attract more traffic than many more items of the same class taken together. In particular, it implies that if there is just one item of a given class, then it will attract more traffic by itself than $M-1$ such items taken together in a corresponding environment where there are $M-1$ such items.

\begin{prop}
Fix two popularity-based search environments $\mathcal{E}$ and  $\mathcal{E}'$ that differ only in the number of items of class 1 ($M_1$ and $M_1'$ respectively). Suppose $M_1 < \frac{M}{1+\beta} < \frac{\beta M}{1+\beta} <  M_1'$, then there exists $\overline{N}$ such that, for any $N \ge \overline{N}$, the total clicking probability ($\rho_{N,M_1}$) by individual $N$ on an item in $M_1$ in environment $\mathcal{E}$  is strictly greater than the total clicking probability ($\rho_{N,M_1'}$) by individual $N$ on an item in $M_1'$ in environment $\mathcal{E}'$, provided $p>0$ is sufficiently small.
\end{prop}

\begin{figure}[!t]
\begin{center}
 \includegraphics[width=.6\columnwidth]{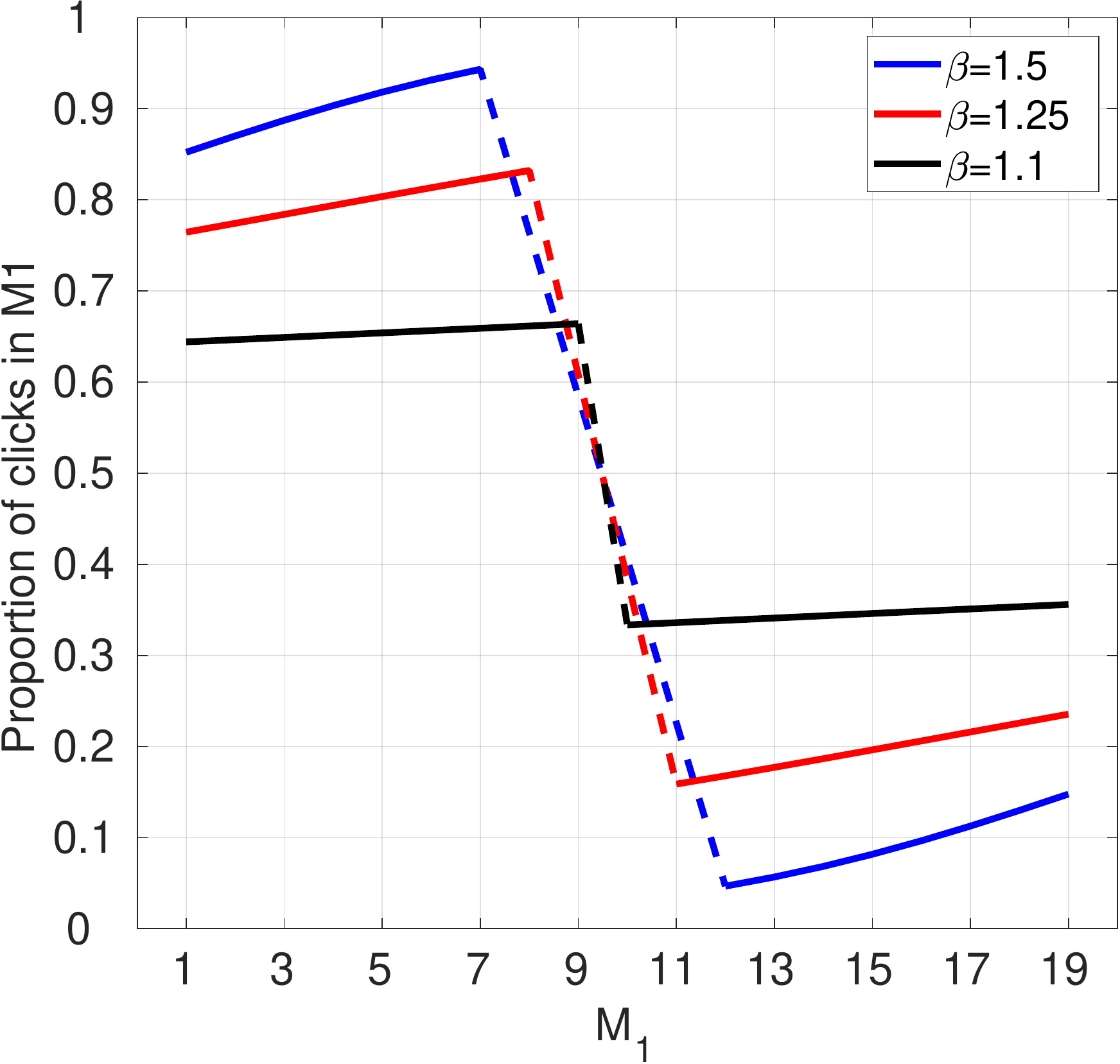}
\end{center}
\caption
{Total traffic on all $M_1$ items of class 1 as a function of $M_1$ for the limit distribution of the process $\rho_n$ of Eq.~\eqref{eq:rho} for different values of $\beta$, with $M=20$, $\Gamma = \{ 0.8, 0.2, 0.5\}$ and $p=p_0=p_1=0.4$.}
\label{fig:fabr}
\end{figure}

The proof is in three steps. First, we characterize a limit ranking ($r_{\infty}$) of the process $\rho_{n}$ defined by Eq.~\eqref{eq:rho} and show it constitutes a (stable) limit. Second, we show it is the unique such limit ranking. Finally, we compute total traffic on all items in $M_1$ at the limit
and show it is over half of total traffic when $M_1 < \frac{M}{1+\beta}$, and hence greater than total traffic on all items in $M_1'$ for $M_1' > \frac{\beta M}{1+\beta}$.

Step 1. Consider $M$ items and a ranking $r_{\infty} \in \{ 1, \ldots, M \}$ defined by $r_{\infty, k} = k$, for $k=1,\ldots,M$. Then, by popularity ranking, $r_{\infty}$ is a (stable) limit ranking of the process $\rho_{n}$ if and only if at $r_{\infty}$: 
\begin{equation} \label{eq:limit}
\rho_{n,1} (r_{\infty}) > \rho_{n,2} (r_{\infty}) > \ldots > \rho_{n,M} (r_{\infty})
\end{equation}
holds for the expected individual $n$.
Suppose $r_{\infty}$ is such that the $M$ items are ranked in two blocks of items of the same class where, if $M_1 < \frac{M}{1+\beta}$,  the first ranked $M_1$ items are all in $M_1$ and the remaining ones (bottom ranked) are all in $M_0$; and symmetrically if $M_1 > \frac{\beta M}{1+\beta}$, the bottom ranked $M_1$ items are all in $M_1$ and the remaining ones (top ranked) are all in $M_0$.  
To see that these constitute limit rankings, suppose $M_1 < \frac{M}{1+\beta}$, and consider the ranking $r_{\infty}$, where the first $M_1$ items are ranked on top. In this case, the corresponding clicking probabilities for $k \in M_1$ (i.e., $k \in \{ 1, \ldots, M_1 \}$) are given by:
\begin{equation} \label{eq:limitclicksM1}
 \rho_{n,k} (r_{\infty}) 
= \frac{\beta^{(M-k)} (1-\gamma_{n})/M_1}{\sum_{k=1}^{M_1} \beta^{(M-k)} (1-\gamma_{n})/M_1 + \sum_{k=M_1+1}^{M} \beta^{(M-k)} \gamma_{n}/M_0},  \nonumber
\end{equation}
and  for $k \in M_0$ (i.e., $k \in \{ M_1+1,\ldots, M \}$) satisfy:
\begin{equation} \label{eq:limitclicksM0}
\rho_{n,k} (r_{\infty}) 
= \frac{\beta^{(M-k)} \gamma_{n}/M_0}{\sum_{k=1}^{M_1} \beta^{(M-k)} (1-\gamma_{n})/M_1 + \sum_{k=M_1+1}^{M} \beta^{(M-k)} \gamma_{n}/M_0}. \nonumber
\end{equation}
Clearly, $\rho_{n,k} (r_{\infty}) > \rho_{n,k+1} (r_{\infty})$ holds within the classes $M_1$ and $M_0$, that is, for $k=1, \ldots, M_1-1$ and for $k=M_1+1, \ldots, M-1$
Hence it suffices to show that $\rho_{n,k=M_1} (r_{\infty}) > \rho_{n,k=M_1+1} (r_{\infty})$. 
This is easily checked for the expected individual $n$ (whose value of $\gamma_n$ is drawn from $\Gamma= \{1,0, \frac{1}{2} \}$ according to the probabilities, respectively, $p_0=p, p_1=p<\frac{1}{2}$ and $p_2=1-2p>0$).

Step 2: To see that the above ranking constitutes a unique limit, we show that no other ranking satisfies Eq.~\eqref{eq:limit} and that, for any other ranking, 
whenever an item is in the less numerous class (say $M_1$ when $M_1 < \frac{M}{1+\beta}$), it will always get strictly more clicks in expectation than the item of the other more numerous class ranked just above. Assume $M_1 < \frac{M}{1+\beta}$, then it can be checked that the two strongest constraints to be satisfied are the ones comparing the clicking probability on the lowest-ranked item of class $M_1$ when $(i)$ it is ranked in the $M$th position while all remaining $M_1-1$ items are ranked in the first $M_1-1$ positions, and when $(ii)$  it is ranked in position $M_1+2$ while all remaining $M_1-1$ items are ranked in the first $M_1-1$ positions. Both are easily seen to be satisfied whenever $p=0$ and $M_1 < \frac{M}{1+\beta}$. By continuity they continue to hold for sufficiently small $p>0$. The case $M_1 > \frac{\beta M}{1+\beta}$, which implies $M_0 < \frac{M}{1+\beta}$ and which has the $M_0$ items ranked on top holds by symmetry.

Step 3. It suffices to show that whenever $M_1 < \frac{M}{1+\beta}$, then the first ranked items in $M_1$ always get strictly more that half the share of the total clicks. Since the share of traffic on the individual items is given by the probabilities $\rho_{n,k} (r_{\infty})$ defined
above, it suffices to show that for any  $M_1 < \frac{M}{1+\beta}$, $\rho_{n,M_1} (r_{\infty})  = \sum_{k=1}^{M_1} \rho_{n,k} (r_{\infty}) > \frac{1}{2}$. Given our assumptions on $p$ and $\Gamma$, this is easily checked for the expected individual $n$ (using the fact that, for any $1 \le K < K' \le M$, $\sum_{k=1}^{K}\beta^{M-K} /K > \sum_{k=1}^{K'} \beta^{M-K'} / K'$). It also implies that when $M_1'>\frac{\beta M}{1+\beta}$ all items in $M_1'$ will (be bottom-ranked) and will obtain less than half the total traffic. $\Box$

\smallskip
When the share of items in $M_1$ is close to one half ($\frac{M}{1+ \beta} < M_1 < \frac{M}{2}$), then there may be multiple limit rankings and the above proof no longer applies. We believe that the few-get-richer result may still go through in these cases, but it is necessary to evaluate the likelihood of the different limit rankings and guarantee that limit rankings are more likely to give classes with fewer items a higher probability of being higher ranked. This proof goes beyond the scope of this paper.
 
Similarly, for interior types ($\Gamma_0 < 1$, $\Gamma_1 > 0$).
As Figure~\ref{fig:fabr} shows for different values of~$\beta$ (see also the simulations in the next section), the effect continues to hold: when $M_1$ has few items (minority case $M_1 < \frac{M}{1+\beta}$) it obtains a larger proportion of clicks than when $M_1$ has many items (majority case $M_1 > \frac{\beta M}{1+\beta}$).


\begin{figure*}[!t]
\begin{center}
\includegraphics[width=.58\columnwidth]{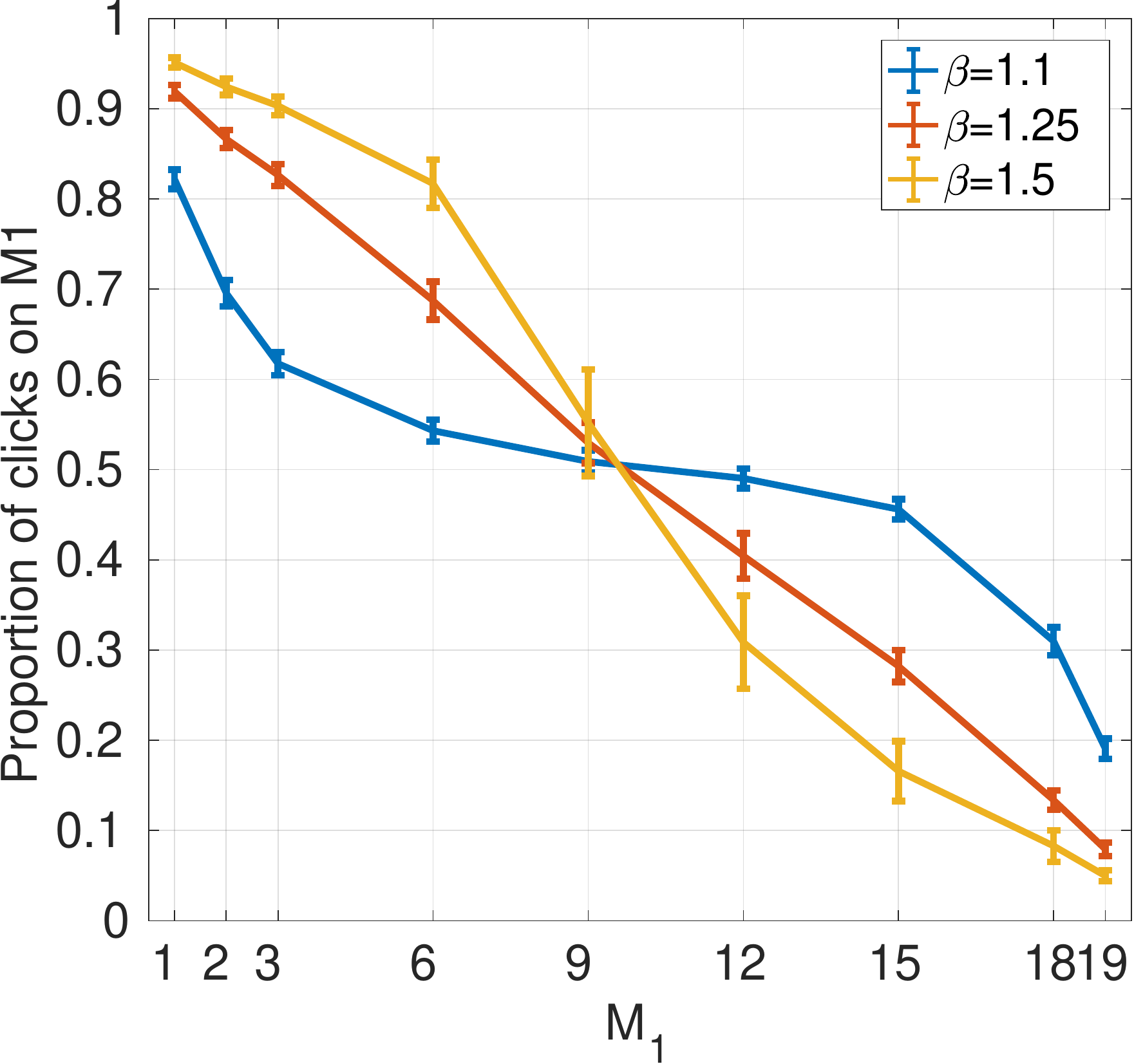}
\includegraphics[width=.58\columnwidth]{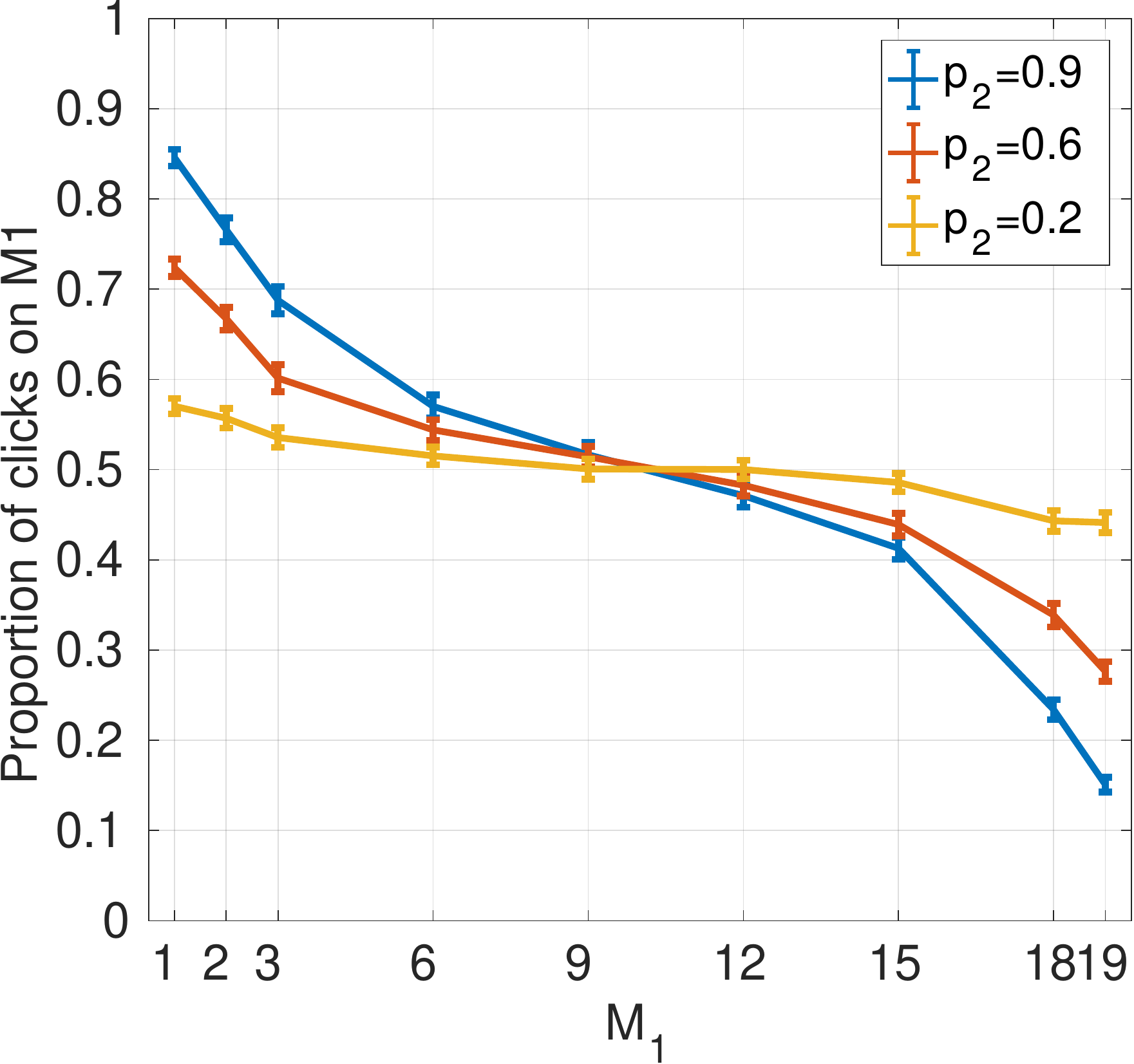}
\includegraphics[width=.58\columnwidth]{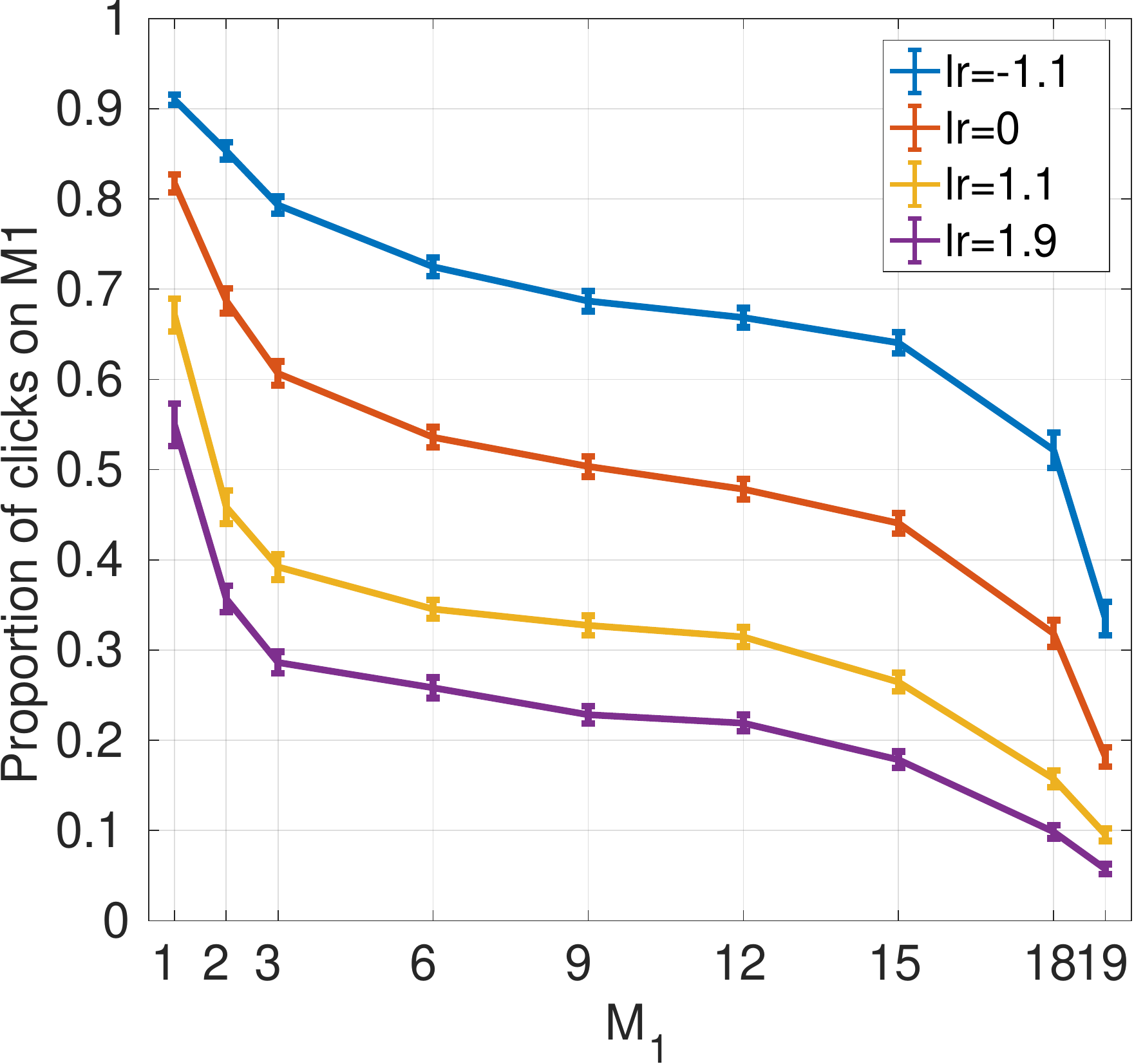}
\end{center}
\caption
{Proportion of clicks on all $M_1$ items of class 1 as a function of $M_1$ for runs of $N=100$, and assuming $M=20$, $\beta=1.1, \Gamma = \{ 0.9, 0.1, 0.5 \}, p_0=p_1=0.4$ and uniform initialization, as baseline, and varying $\beta$ (left plot), $p_2=1-p_0-p_1$ (middle plot), and the log-ratio $\log{\frac{p_0}{p_1}}$ (right plot). Error bars indicate confidence intervals for 100 different random realizations.}
\label{fig:clicks}
\end{figure*}

\section{Simulations of the Model}
\label{sec:sims}

We now analyze through simulations the presence of the few-get-richer effect in different settings.
Our focus is on analyzing the click-through rate (CTR), defined as the the ratio of the probability of clicking on an item in $M_1$ to the total number of clicks $N$ (which we fix at $N=100$ users), as a function of $M_1$ for different settings. 
We mainly consider stochastic choices with $\Gamma = \{ 0.9, 0.1, 0.5 \}$, which means that non-indifferent users do not exclusively click on one of the two classes.
We assume that the ranking $r_{n}$ is proportional to the number of clicks that the item received at time $n$.
As before, we consider a ranking of $M=20$ items with the $M_1$ items initially at the bottom. In this case, instead of the limit ranking, we characterize the dynamic transient during which the minority class may reach the top of the ranking. This makes our analysis dependent on the initial conditions. For an item $m$ initially at position $r_{1,m}$, we assume a \emph{uniform} initialization, with all items having one click.

\subsection{Dependence on $\beta$}
We first consider the dependence on the ranking effect, parametrized by $\beta$.
Larger values of $\beta$ correspond to a stronger relative ranking effect compared to the propensities.
Here, we consider a symmetric case with $p_0=p_1=p=0.4$.


Figure~\ref{fig:clicks} (left) shows the CTR for different values of $\beta$. 
The resulting CTR is almost symmetric.
We observe a monotonic decrease as a function of $M_1$ in all cases, showing, in particular, that the minority class \emph{always} receives more clicks than the majority one.

We can differentiate between three cases, corresponding to small, intermediate and large values of $\beta$, respectively.
In general, larger values of $\beta$ lead to a relatively larger CTR for the minority class, due to relatively larger probability of clicking on top ranked items.
For small values of $\beta$, the CTR is the smallest, but even here the effect as a function of $M_1$ is particularly pronounced for small (and large) $M_1$.
For intermediate values of $\beta$, the flat region for intermediate values of $M_1$ disappears and the CTR decreases monotonically with constant slope, indicating a decrease independent of $M_1$.
Finally, for larger values of $\beta$, the CTR is largest, and the effect as a function of $M_1$ is smallest for small (and large) $M_1$.
In this extreme case, for $M_1=1$, the single item at the bottom quickly reaches the top and attracts almost all the traffic, leaving almost no traffic to the remaining $M-1$ items of the other class.

From these simulations, we can conclude that 
the few-get-richer effect is robust to varying $\beta$.



\subsection{Dependence on $p_2$}


\begin{figure*}[!t]
\begin{center}
\includegraphics[width=.6\columnwidth]{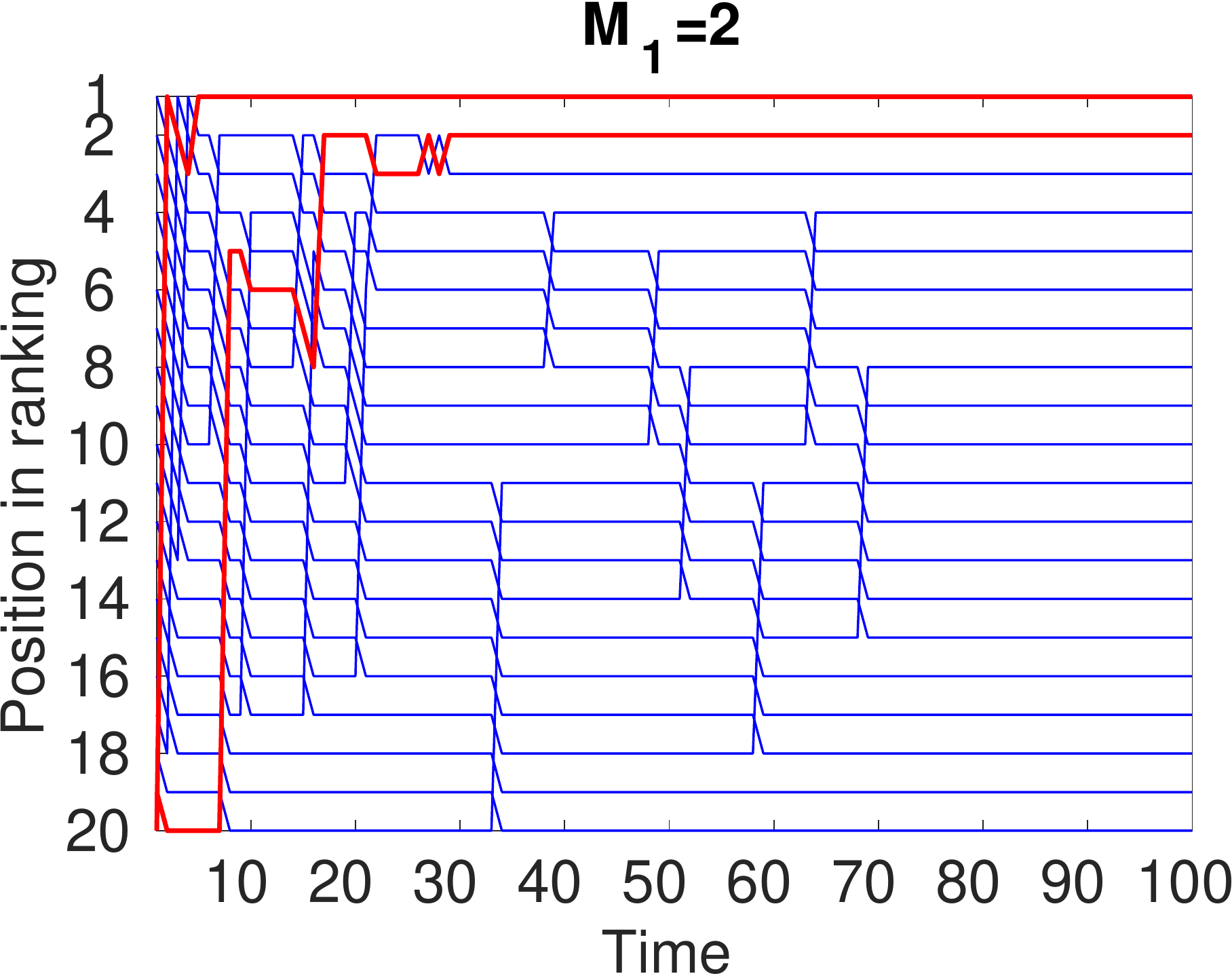}
\includegraphics[width=.6\columnwidth]{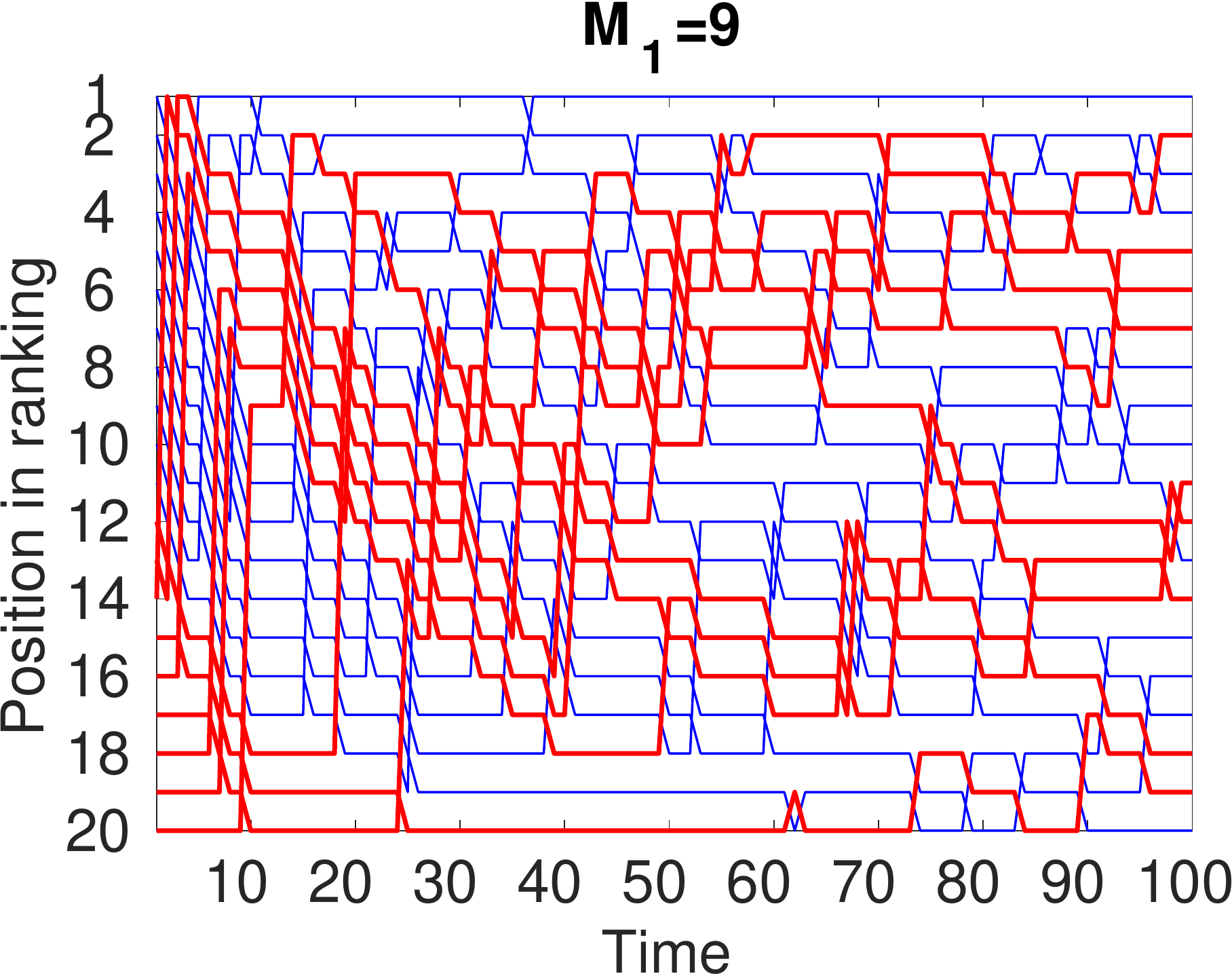}
\includegraphics[width=.6\columnwidth]{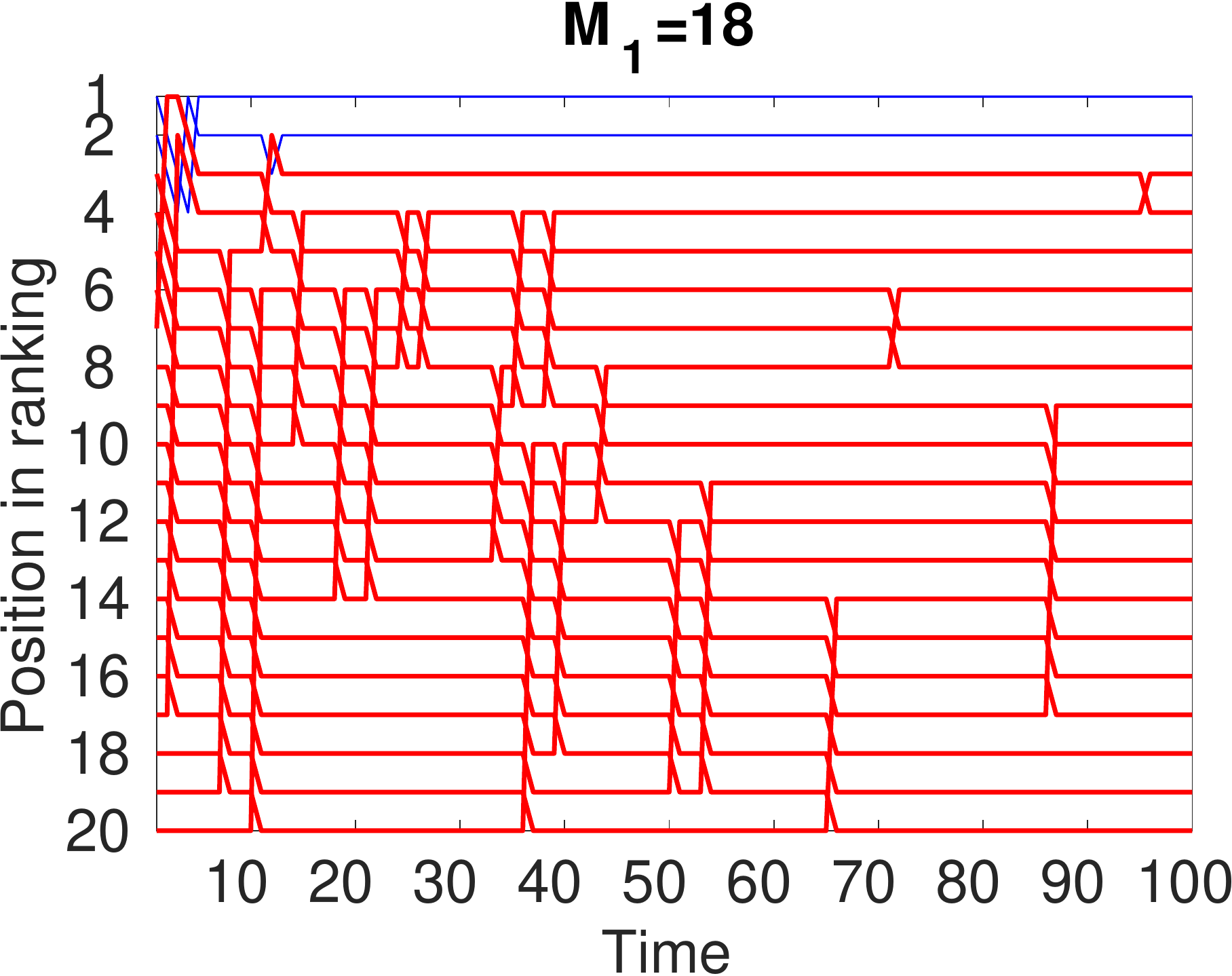}
\end{center}
\caption
{Examples of ranking evolution for different values of $M_1$. 
The items in $M_1$ always start at the bottom of the ranking.
With $M_1=2$ (left plot), the two items quickly move to the top.
With $M_1=9$ (middle plot), the items in $M_1$ are spread through the different ranking positions.
Finally, with $M_1=18$ (or $M_0=2$) (right plot), the items $M_1$ eventually stay at the bottom.}
\label{fig:rankings}
\end{figure*}

We now analyze the CTR as a function of the proportion of indifferent users $p_2=1-p_0-p_1$.
For this, we choose $\beta=1.1$ and vary the proportion of indifferent users while keeping $p_0=p_1$.
To better analyze this dependence, we consider extreme preferences, so that only indifferent users can really click on both items, that is, we assume $\Gamma=\{1,0,0.5\}$.

Figure~\ref{fig:clicks}(middle) shows the results. We observe that, in this case, the few-get-richer effect is more pronounced as $p_2$ increases.
Having a larger proportion of indifferent users results in relatively more clicks per item, and hence a larger amplification of ranking effect, which is a key ingredient for the few-get-richer effect to emerge. With less extreme preferences, e.g., $\Gamma=\{0.9,0.1,0.5\}$, the effect in $p_2$ is still present, but less pronounced (data not shown).

We conclude that the indifferent users play a key role in amplifying the effect of the ranking, and that in general, having a larger proportion of them contributes importantly towards the few-get-richer effect.


\subsection{Dependence on the ratio $\frac{p_0}{p_1}$}

So far we have considered cases where the distribution of user types was symmetric, $p_0=p_1=p$.
In practice this need not be the case since, e.g., the minority class might also be preferred by a minority of users. 
We analyze the effect as a function of $lr=\log\frac{p_0}{p_1}$. 


Figure~\ref{fig:clicks}(right) shows the results for different values of $lr$. 
As expected, the few-get-richer effect is more pronounced when there is a larger relative proportion of users that prefer the minority item.
To see this, compare the blue line, where $p_1=0.6>p_0=0.2$ (there are three times more users who prefer the `minority' item), and the purple line, where $p_1=0.1<p_0=0.7$.
We see that the effect is still present for this choice of parameters, even when there are seven times more users who prefer the majority item. Here, the proportion of indifferent users is set to $p_2=0.2$. Consistent with the results of the previous subsection, if we increase $p_2$ and keep the same ratios, the effect becomes more pronounced.
We conclude that the effect is also robust to different ratios of proportions $lr$.

\subsection{Ranking evolution}
Finally, we illustrate the typical behavior of the ranking evolution for different values of $M_1$, assuming our usual baseline parameter values for $\beta$, $\Gamma$, and $p$ and uniform initialization.
Figure~\ref{fig:rankings} shows the results, which confirm the idea that minority items tend to move towards the top.

\section{Online Experiment with Human Participants}\label{sec:exp}

To test the predictions of the model, we executed an online experiment in which participants clicked on one out of 20 possible options (pictures).\footnote{Data from the experiment are available at \url{https://osf.io/nwjyf/}.} The options belonged to two classes, $M_0$ and $M_1$, just as in the model. As in the simulations reported above, we measure the effect of the popularity-based ranking of options as the total number of clicks attracted by items of type $M_1$. To this end, we created 8 independent conditions. In conditions D1 to D4, the ordering of the options changed dynamically as a function of the number of clicks received by each option, with the most popular option (in terms of cumulative number of clicks) at the top of the screen and the least popular option at the bottom of the screen. The conditions differed in terms of the number of options in $M_0$ and $M_1$, as summarized in Table~\ref{tab: Human Experiment}. The number of options in $M_1$ was 17, 12, 8, or 3 in conditions D1, D2, D3 and D4, respectively.

These four `dynamic' conditions were matched to four `static' conditions with the same sets of options $M_0$ and $M_1$. In the static conditions, the ranking of options did not change over time and was set to the initial ranking in the matching dynamic condition. We denote these static conditions by S1, S2, S3, and S4.

Comparisons of matched pairs of conditions allows us to establish the causal effect of the ranking algorithm on the total traffic attracted by each option. Comparisons of the 4 dynamic conditions allows us to test the prediction that the total traffic attracted by options in $M_1$ decreases with the number of options in $M_1$.

To keep the experimental setup as simple as possible, the options in our experiment were not news sources, but pictures of dogs and cats. In order to activate their preferences for cat or dog pictures, participants were first asked whether they were a `cat person', a `dog person', or `neither a cat person nor a dog person.' Answers to this identity question allowed us to compute the proportion of participants in each of the 3 types discussed in the model section. Then participants were shown a set of 20 pictures in a vertical list. In this setup, $M_0$ is the set of cat pictures (initially at the top of the screen) and $M_1$ is the set of dog pictures (initially at the bottom). The initial popularity of all pictures was set to 1, consistent with the uniform initialization of the previous section.

\subsection{Methodological Details}
We recruited 786 participants on Amazon Mechanical Turk. It was administered via a Qualtrics survey embedded in the Amazon Mechanical Turk webpage as an iframe. After signing up for the task, participants read the informed-consent form. Then they were randomly assigned to one of 8 conditions. They first answered a question about their type: ``Are you more of a cat person or a dog person?'' with three possible choices ``I am a cat person'' / ``I am neither a cat person nor a dog person'' / ``I am a dog person.'' On the next screen they were shown 20 buttons with  ``Please click on a photo from the following list of photos of cats and dogs and rate it according to your liking.'' The buttons were displayed in a vertical list. Participants could initially see 3 to 4 buttons and had to scroll down to access the other buttons. 
After clicking a button, participants were shown the corresponding picture and gave it a rating of 1 to 5 stars. The rating task was presented as a reason to ask participants to select an item according to their preferences. The collected ratings are not discussed because our interest is only in the clicking behavior of the participant. Participants were paid \$0.15 for their time.


\subsection{Results}

\begin{table}
\caption{Experiment results. The `Sim1' rows report the average traffic attracted by Dog pictures ($M_1$) over 1000 simulations of the choice model with a setting matching the exact number of participants of each identity type in each condition. The `Sim2' rows report the average traffic attracted by Dog pictures ($M_1$) over 1000 simulations of the choice model with 100 users in each condition where the numbers of users who are a `dog person', `neither a dog person nor a cat person' and a `cat person' are 55, 15 and 30, respectively (same frequencies for all conditions).\label{tab: Human Experiment}}

\begin{tabular}{cc|c|c|c|c}
\multicolumn{6}{c}{}\tabularnewline
\hline 
\hline 
\multicolumn{2}{c|}{\# Cats ($M_{0}$)} & 3 & 8 & 12 & 17\tabularnewline
\hline 
\multicolumn{2}{c|}{\# Dogs ($M_{1})$} & 17 & 12 & 8 & 3\tabularnewline
\hline 
\hline 
\multicolumn{6}{c}{Dynamic Button Ordering}\tabularnewline
\hline 
\multicolumn{2}{c|}{Condition} & D1 & D2 & D3 & D4\tabularnewline
\hline 
\multicolumn{2}{c|}{\# participants} & 96 & 102 & 99 & 101\tabularnewline
\hline 
\# participants  & Cat person & 34 & 30 & 24 & 29\tabularnewline
\cline{2-6} 
in each & Neither & 9 & 21 & 11 & 16\tabularnewline
\cline{2-6} 
type & Dog person & 53 & 51 & 64 & 56\tabularnewline
\hline 
Dog  & Experiment & .53 & .69 & .76 & .71\tabularnewline
\cline{2-6} 
traffic  & Sim1 & .46 & .56 & .73 & .76\tabularnewline
\cline{2-6} 
share & Sim2 & .47 & .60 & .67 & .75\tabularnewline
\hline 
\hline 
\multicolumn{6}{c}{Static Button Ordering}\tabularnewline
\hline 
\multicolumn{2}{c|}{Condition} & S1 & S2 & S3 & S4\tabularnewline
\hline 
\multicolumn{2}{c|}{\# participants} & 96 & 101 & 95 & 96\tabularnewline
\hline 
\# participants & Cat person & 34 & 30 & 25 & 33\tabularnewline
\cline{2-6} 
in each & Neither & 13 & 19 & 9 & 15\tabularnewline
\cline{2-6} 
type & Dog person & 49 & 52 & 61 & 48\tabularnewline
\hline 
Dog  & Experiment & .44 & .37 & .40 & .27\tabularnewline
\cline{2-6} 
traffic  & Sim1 & .41 & .37 & .39 & .28\tabularnewline
\cline{2-6} 
share & Sim2 & .44 & .39 & .35 & .30\tabularnewline
\hline 
\hline 
 & \multicolumn{1}{c}{} & \multicolumn{1}{c}{} & \multicolumn{1}{c}{} & \multicolumn{1}{c}{} & \tabularnewline
\end{tabular}

\end{table}



\subsubsection{Participant Types}
30\% indicated they were `Cat persons' ($p_0=0.30$), 55\% `Dog persons' ($p_1=0.55$), 15\% neither ($p_2=0.15$).
\subsubsection{Total Traffic Attracted by $M_1$ (Dog pictures)}
Dog pictures were initially ranked at the bottom the screen. First we discuss the effect of popularity-based ranking of options on the total traffic attracted by Dog pictures. In the 4 dynamic conditions, Dog pictures attracted substantially more traffic than in the corresponding static condition (compare the two rows `Experiment' in the top and bottom panels of Table~\ref{tab: Human Experiment}). In other words, ordering the options in terms of popularity had a systematic effect on the share of traffic attracted by options that started at the bottom of the choice screen.

Unsurprisingly, in all static conditions, the total traffic attracted by Dog pictures was lower than 50\%. The set of Dog pictures attracted more traffic when there were relatively more Dog pictures. For example, while 17 Dog pictures attracted 44\% of the traffic, 3 Dog pictures attracted 27\%.

The pattern is completely different in the dynamic conditions. First, in all dynamic conditions, the total traffic attracted by Dog pictures was higher than 50\%. The most important finding is that the total traffic attracted by Dog pictures was \emph{larger} with just 3 Dog pictures (and 17 Cat pictures) than with 17 Dog pictures (and 3 Cat pictures)!  Similarly, the total traffic attracted by Dog pictures was \emph{larger} with 8 Dog pictures (and 12 Cat pictures) than with 12 Dog pictures (and 8 Cat pictures). These results are consistent with the predictions of our model. The total traffic attracted by Dog pictures did not decrease monotonically with the size of the set of Dog pictures (it is larger with 8 Dog pictures than with 3 Dog pictures). This pattern is seemingly inconsistent with the predictions of the simulations (Fig.~\ref{fig:clicks}). Yet, this lack of monotonicity can be explained by the fact that the conditions were not perfectly balanced with respect to the distribution of identity types of the participants. In condition D3 (8 Dog pictures) there were substantially more Dog lovers (64/99) than in condition D4 (3 Dog pictures, 56/101). This is a chance event resulting from the fact that we elicited the identity type of the participants after the random assignment into conditions, thus we could not balance types within conditions. 

\subsubsection{Estimated Parameters of the Choice Model} 
We estimated the parameters of the stochastic choice model, Eq.~\eqref{eq:rho}, using maximum likelihood on the whole dataset. Our parameter estimates are $\beta = 1.22$, $\Gamma_C=.74$ for Cat persons ($\Gamma_0$ in the model), and $\Gamma_D=.08$ for Dog persons ($\Gamma_1$ in the model). Cat and Dog persons had strong tendencies to choose pictures from the class consistent with their identity type.

We used these estimated parameters to simulate choices and the dynamics of picture ranks in the 8 experimental conditions. Simulations reported in rows `Sim1' in Table~\ref{tab: Human Experiment} indicate a close match between the simulated data and the actual traffic proportions attracted by Dog pictures (both in static and dynamics conditions). 

We also used the estimated parameters to simulate what would happen if there were 100 participants in each condition and if the distributions of identity types were the same in all conditions (`Sim2' rows). We find a decreasing monotonous relation between the number of Dog pictures and the share of traffic attracted by Dog pictures. This pattern is consistent with the qualitative prediction of our model.

\subsubsection{Summary} 
Overall, the results indicate that the experimental setting falls within the boundary condition of application of our theory. They provide a proof-of-concept that popularity-based ordering of options can lead to the counter-intuitive phenomenon that when there are fewer options of one class, the total share of traffic attracted by this class of options becomes larger.

\section{Discussion \& Conclusion}\label{sec:conc}
The few-get-richer effect can have both positive and negative effects on the quality of the information people obtain from search results. On the positive side, when there are few relevant items, the few-get-richer effect may help them become top-ranked, making them more accessible. At the same time, the few-get-richer effect can contribute to the spread of misinformation. It may help few irrelevant or `fake news' items become top-ranked, especially if there is a strong preference for such items and only few websites report them. 

Our analyses highlight a potential unintended effect of regulations that `ban' particular `alternative' news sources known for spreading `fake news'. When a sizeable proportion of users have a preference for identifiable `alternative' news sources, the removal of some of these news sources might lead to an increase in the total traffic attracted by the remaining `alternative' news sources. This could result in more overall exposure to `fake news'!

To neutralize the few-get-richer effect, our theory suggests that it may be advisable to keep track of the number of items in each class when incorporating clicks in the search engine algorithm. Ideally, the ranking algorithm should use the popularity of the different items in a way that is neutral to the number of items in each class.



The few-get-richer effect also has implications for the design of recommender systems. The learning efficiency of these systems is impeded by the \emph{presentation bias} problem: items shown to the user can get clicks whereas items not shown get no clicks. The  recommender system thus cannot learn about the relevance of the latter items. A popular solution to this challenge is the \emph{explore-and-exploit} approach, in which some items from a minority class are randomly inserted in the search results ~\cite{agarwal2009explore}. This purportedly increases the amount of exploration (clicks on the minority class), and thus increases the learning opportunities of the system at the cost of slightly hurting the user experience~\cite{mcnee2006being, herlocker2004evaluating}. The few-get-richer effect suggests precisely the opposite. Adding more of those items might reduce, rather than increase, the total amount of exploration.

\bibliographystyle{ACM-Reference-Format}
\bibliography{bibliography}

\end{document}